\documentclass[english,aps,prper,reprint,showpacs,titlepage,longbibliography]{revtex4-2}   

\usepackage{soul}
\usepackage{xcolor}
\soulregister{\textit}{1}

\definecolor{lightgray}{gray}{0.9}

\sethlcolor{lightgray}
\usepackage[utf8]{inputenc}
\usepackage[compatibility=false]{caption} 
\usepackage{booktabs} 
\usepackage{tabularx} 
\usepackage{ragged2e} 
\usepackage[T1]{fontenc}	
\usepackage{geometry}
\usepackage{times}
\usepackage{hyperref}  
\hypersetup{colorlinks=true,urlcolor=blue,citecolor=blue,linkcolor=blue} 
\usepackage{array}
\usepackage{enumerate}
\usepackage{enumitem}
\usepackage{amsmath}
\usepackage{caption}
\usepackage{amssymb}
\usepackage{tikz}
\usepackage{graphicx}
\usepackage{multirow}
\usepackage{tcolorbox}

\usepackage{float}
\usepackage[utf8]{inputenc}

\usepackage[normalem]{ulem}
\usepackage{setspace} 
\usepackage{placeins} 
\usepackage{adjustbox} 
\usepackage{graphicx}
\usepackage{subcaption}
\usepackage[utf8]{inputenc}
\usepackage{booktabs}
\usepackage{array}
\usepackage{geometry}
\usepackage{lipsum} 

\newcolumntype{C}[1]{>{\centering\arraybackslash}p{#1}} 
\newcolumntype{L}[1]{>{\raggedright\arraybackslash}p{#1}}

\begin{document}
\begin{titlepage}

\title{Assessing Physics Students' Scientific Argumentation using Natural Language Processing }

 \author{Winter Allen}
 \affiliation{Dept. of Physics and Astronomy, Purdue University, 525 Northwestern Ave, West Lafayette, IN 47907, U.S.A.} 

 \author{Carina M. Rebello}
 \affiliation{Dept. of Physics and Astronomy, Purdue University, West Lafayette, IN 47907, U.S.A.}
  
 \author{N. Sanjay Rebello}
 \affiliation{Dept. of Physics and Astronomy / Dept. of Curriculum \& Instruction, Purdue University, West Lafayette, IN 47907, U.S.A.} 


\begin{abstract}
Scientific argumentation is a core science and engineering practice and a necessary 21st Century workforce skill. Due to the nature of large enrollment classes, it is difficult to individually assess students and provide feedback on their scientific argumentation. The recent developments in Natural Language Processing (NLP) and Machine Learning (ML) provide new opportunities to analyze large collections of student writing efficiently. In this study, we investigate how undergraduate students' scientific argumentation evolves across four semesters of an introductory calculus-based physics course as increasingly structured argumentation scaffolds were introduced. We investigate the  use of NLP and ML, specifically topic modeling, to analyze student scientific argumentation across those semesters. We report on the emergent themes present in each semester. Our findings show a clear shift in the thematic focus of student arguments corresponding to the level of scaffolding provided. In semesters with minimal scaffolding, students' arguments emphasized procedural and surface-level features, while semesters with explicit scaffolds exhibited greater concentration around physics-principle-based themes. These results suggest that structured scaffolding supports students in constructing more conceptually grounded scientific arguments and highlights the potential of NLP and ML as scalable approaches for evaluate broad trends in students' scientific argumentation. 

    \clearpage
  \end{abstract}

\maketitle
\end{titlepage}
\maketitle
\section{Introduction}
We live in an age where misinformation is spread with a click of a button. We are constantly bombarded with dissenting information requiring strong critical thinking skills to wade through to the truth. Students often view science as another subject to memorize and regurgitate facts, rather than as a dynamic, evolving body of knowledge. This fundamental issue can be addressed by encouraging students to think critically about their work and engage in scientific argumentation in a classroom. 

Student problem solving has been a relevant and rich topic in Physics Education Research (PER) for many years \cite{doktor2014, Tulminaro2007, van1991learning}. Students often prioritize memorizing final answers over developing a deeper understanding of the problem-solving process \cite{Tulminaro2007, dufresne1997solving}. To foster growth in problem-solving, it is crucial not only to understand why students solve problems the way they do but also to help them reflect on their own problem-solving strategies, allowing them to develop as both learners and future scientists. 

Several research-based strategies have attempted to address the difficulties that students have with problem solving \cite{doktor2014}. Research has shown that scientific argumentation can improve problem solving \cite{NUSSBAUM2003384}. This idea is deeply rooted in philosophy \cite{Toulmin1958-TOUTUO-6} and has evolved significantly through educational research \cite{Toulmin1958-TOUTUO-6, mcneill2010scientific}. Scientific argumentation is a proven strategy to help improve critical thinking that provides a schema for justifying the relevance of the retrieved knowledge in problem solving \cite{asterhan2007effects, mcdonald2010influence, nussbaum2003argument}. To construct an argument students must justify their methods and decisions as they solved a problem, go through every step they took up to their solution, and provide evidence and reasoning for their process. In the context of problem-solving in physics, scientific argumentation involves not only an explanation of conceptual knowledge and methods but the ability to justify reasoning with empirical evidence and logical consistency. \cite{cho2002effects} 

Within PER, scientific argumentation has been shown to enhance students' ability to link theoretical knowledge with practical problem-solving skills \cite{rebello2019using}. This process encourages students to think critically about the methods they use and the evidence they gather, promoting skills that are essential for expert-like problem solving. Peer argumentation in physics classrooms also fosters collaborative learning, where students refine their ideas through group discussions and critique, further advancing their conceptual understanding and reasoning abilities \cite{berland2011classroom}. The iterative nature of scientific argumentation i.e. the process of reflecting and evaluating one's solution -- aligns with the goals of promoting both content mastery and the development of scientific critical thinking.

Scientific argumentation can be studied in both oral and written modalities \cite{hakyolu2016interplay}.  In this work, we focus on investigating scientific argumentation in written essays in which students describe their strategies for solving problems. One of the major challenges that we face is systematically teaching and assessing student written essays in large enrollment classes due to the prohibitive time it takes for instructors or teaching assistants to read students' written work and gauge their argumentation quality. Recent developments in Large Language Models (LLMs), Natural Language Processing (NLP), and Machine Learning (ML) may afford us the opportunity to address these challenges. ML is a subset of Artificial Intelligence (AI) that is the development of algorithms to detect patterns in datasets. NLP and LLM's are subsets within ML that focus on and aid in the understanding of human language. Recently, there have been many studies on using LLMs, NLP, and ML in assessing student writing \cite{allen2025studentsperceptionslargelanguage, bralin2023analysis, sirnoorkar2025feedbackclicksintroductoryphysics, savage2025using}. \cite{anand2023sciphyrag, wulff2024physics, polverini2023understanding, kieser2024using, tschisgale2023}. Researchers have used these tools, such as supervised and unsupervised NLP, to assess student strategy essays \cite{munsell2021, tschisgale2023} and explore the viability of utilizing LLMs in physics education \cite{kieser2024using, polverini2023understanding, anand2023sciphyrag,wulff2024physics}. NLP and ML provide unique tools for analyzing large amounts of student text, allowing for the identification of patterns while still enabling qualitative, in-depth studies.


We aim to address the gap in literature on longitudinal studies that investigate how students' scientific argumentation, in the context of physics problem-solving, develops over time in response to the use of research-based instructional strategies to improve scientific argumentation. Due to the large datasets from large enrollment classes, it becomes increasingly difficult to track global student progress on argumentation in response to scaffolding. However, with the advent of ML, it is possible to leverage classic unsupervised machine learning (UML) techniques to study the longitudinal development of scientific argumentation skills for large numbers of students.

To approach assessing students' argumentation with ML in the context of problem-solving we have designed a study that introduces scientific argumentation through a series of scaffolds in the recitation portion of an introductory physics course. We offer increasing  levels of scaffolds through a four semester study. We then utilize unsupervised machine learning techniques to assess how student arguments change through the semesters.

Our research question (RQ) is:



\vspace{0.2cm}


\indent
\begin{minipage}{0.90\linewidth}
    \textit{ How does the quality of students' scientific argumentation in the context of problem-solving in physics, evolve in response to scaffolding and fading over multiple weeks of instruction?}
    \label{RQ1}
\end{minipage}
\\

 The current article is structured as follows: in the next section, we provide the background on prompt scientific argumentation and unsupervised machine learning. In Section III, we describe the study’s data, context, scaffolding, and analysis. In the following section, we present the results and discussion. Finally, in Section V, we then conclude with study’s implications, limitations, and directions for future research.

\section{Background} 

\subsection{Scientific Argumentation}
It has been shown that by constructing explanations students may change their view of the nature of science \cite{bell2000scientific}.  As scientists, we know it is important to critically reflect on our work, whether it be in research or solving a simple freshman level problem. We not only need to understand what decisions we made and why we made them in the problem-solving process, but we need to be able to present a well supported argument in support of our process. Research suggests that students tend to struggle with developing scientific arguments \cite{berland2009making, kuhn1993science}, especially with finding appropriate evidence and constructing their reasoning \cite{berland2010learning,kuhn2010teaching} and distinguishing between various elements of an argument \cite{forman1998you,jimenez2000doing}.  

Despite these difficulties, engaging in scientific argumentation has shown to be beneficial to students. Not only does it aid in their own understanding of the problem, but it helps them learn to communicate and support their own findings \cite{zohar2002fostering}. To aid students in constructing scientific arguments, argumentation scaffolds can elicit students’ participation in scientific argumentation \cite{jonassen2010arguing} and a conducive learning environment can support students to solve problems, compare solutions, consider alternatives, and justify choices \cite{berland2010learning}, \cite{jimenez2000doing}. Appropriate scaffolds include justification prompts \cite{xun2004conceptual} and question prompts \cite{cho2002effects}in instructional materials that help students articulate the rationale for their problem-solving steps and urge them to reason using evidence and justifications \cite{christodoulou2014science,mcneill2010scientific} based on underlying principles \cite{schworm2007learning}. However, most undergraduate physics courses do not facilitate scientific argumentation. Curricula that facilitate more expert-like problem solving can positively influence students’ epistemic beliefs and expectations around problem solving \cite{wampler2013relationship}. In more recent work, Rebello et al. \cite{rebello2019scaffolding,rebello2019using} found positive effects of using scientific argumentation in physics courses for future elementary teachers as well as future engineers. 

In recent years, scientific argumentation has been studied in various science disciplines including biology \cite{anwar2019analyzing} \cite{dorfner2018biology}, chemistry \cite{erduran2019argumentation},  and physics \cite{yang2023effects}. There are multiple ways to assess scientific argumentation \cite{abi2011perceptions} that rely on context in which the argument is presented, nature of the task being performed, and the specific learning goals \cite{sampson2008assessment}. For example, scientific argumentation in inquiry-based tasks may be assessed differently than in conceptual problem-solving activities. Moreover, these assessments must account for the complexity of scientific reasoning, the use of evidence, and how students articulate and justify their claims \cite{crujeiras2017high}. As a result, educators and researchers have developed diverse methodologies, such as Toulmin’s argumentation model \cite{toulmin2003uses,Toulmin1958-TOUTUO-6}, which emphasize the structure of arguments, as well as rubrics that measure the quality of evidence and reasoning in students' responses \cite{berland2016epistemologies}. 

Toulmin, a British philosopher, proposed breaking down arguments into six components: claim, grounds, warrant, qualifier, rebuttal, and backing. \cite{Toulmin1958-TOUTUO-6} A claim is the base purpose of an argument. The grounds are the evidence of the argument that support the claim. The warrant links the grounds to the claims. This can be explicitly stated or implied. The qualifier is used in wording the claim, while the rebuttal and backing are implied. The rebuttal acknowledges things that may contradict the claim. The backing establishes the relevance of the warrant.  \cite{toulmin2003uses}, \cite{karbach1987using} Toulmin's model is based on arguments present primarily in law; however, this model has been particularly useful in examining how students construct arguments and justify their claims.

McNeill and Krajcik \cite{mcneill2011supporting} further adapted Toulmin’s model for use in science education by simplifying it into the Claim-Evidence-Reasoning (CER) framework. In their framework, the claim is an assertion or conclusion about a phenomenon, the evidence consists of scientific data supporting the claim, and the reasoning explains the relevance of the evidence. CER has become a popular framework in K-12 education, where students are encouraged to construct arguments using data to support their claims \cite{mcneill2008inquiry, wang2020scrutinising}. Given the effectiveness of CER in K-12, there is a strong rationale for exploring its adaptation in undergraduate physics education, where developing students' ability to argue scientifically can enhance their problem-solving and critical thinking skills.

Scientific argumentation has been a rich topic to study in science education in K-12 classrooms in recent years. Erduran and Park \cite{erduran2023argumentation} performed a search for manuscripts focusing on argumentation in K-12 contexts between 2003 and April 2022. During this period, they identified 13 published studies that explored scientific argumentation in secondary physics classrooms. These studies have provided critical insights into how students develop argumentation skills at the K-12 level, emphasizing the importance of scaffolding and structured support for helping students engage with complex scientific reasoning. However, despite the growing interest in this area, the relatively small number of studies over two decades suggests that research on argumentation, particularly in physics, is still in its early stages. Furthermore, these findings highlight the need for continued efforts to expand scientific argumentation research, particularly given the increasing emphasis on argumentation as a critical component of scientific literacy in modern curricula \cite{berland2016epistemologies}.

Even less research has been completed on scientific argumentation in undergraduate physics classrooms. While Erduran and Park \cite{erduran2023argumentation} found 13 manuscripts on scientific argumentation in secondary settings from 2003 to 2022, only nine manuscripts were published during the same period focusing on tertiary settings, highlighting the gap in research at the undergraduate level. In these nine manuscripts, there was a range of focus. The lack of study in this context does raise concerns as argumentation plays a critical role in developing advanced scientific reasoning and expert-like critical thinking skills, which are essential for physics students at the tertiary level as they transition into more complex problem-solving tasks that will aid them in the 21st century job market. 

In these nine manuscripts, a range of focus areas emerged. Some research concentrated on the epistemic tools learners use to construct scientific arguments \cite{cikmaz2021examining}, investigating how students draw upon their knowledge, reasoning, and evidence to build coherent, well-supported arguments. Other studies examined the relationship between student content knowledge and their performance on argumentation tasks \cite{jonsson2016student}, showing that students’ ability to engage in scientific argumentation often hinges on their depth of understanding of the underlying physical principles. Additionally, some research has explored the ways in which students use arguments across different disciplinary contexts, such as mathematics and physics, highlighting the challenges students face when applying argumentation strategies in varied subject areas \cite{van2019translating}.

Despite these valuable insights, the limited volume of research on scientific argumentation in undergraduate physics classrooms points to the need for continued research in the topic. There is a lack of comprehensive, longitudinal studies that examine how scientific argumentation can be systematically integrated into traditional physics curricula. Most undergraduate physics courses emphasize procedural problem-solving without incorporating active retrieval practices or argumentation-based activities that strengthen long-term memory and conceptual understanding. As a result, students have limited opportunities to engage in the kind of reflective, evidence-based reasoning that scientific argumentation requires. We aim to contribute to the existing literature by examining how scientific argumentation can be effectively integrated into undergraduate physics recitations to improve the quality of students’ written arguments. Further, given the large dataset of our investigation, we will employ systematic machine learning methods to track these changes across large textual datasets.

\subsection{Machine Learning}

Machine learning (ML) is a branch of artificial intelligence (AI) that focuses on the development of statistical algorithms to understand patterns in unseen data. ML includes both unsupervised and supervised \cite{sarker2021machine} methods that are summarized below.

Supervised ML relies on known, labeled data to train an algorithm to predict patterns in unseen data. In contrast, unsupervised ML focuses on discovering hidden patterns within datasets without explicit human guidance \cite{sarker2021machine}. Some popular examples of machine learning are clustering (i.e. K-Means \cite{macqueen1967multivariate}, DBSCAN \cite{ester1996density}, HDBSCAN \cite{campello2013density} and topic modeling (i.e. LDA \cite{blei2003latent}, LSA \cite{landauer1997solution}, and NMF\cite{lee1999learning}). 

In this study, we used unsupervised machine learning to analyze our data. This decision is motivated by several factors. First, the large sizer of  our data set makes it prohibitively time consuming to qualitatively extract emergent themes in a way that is reproducible and reliable. Second, although we have expectations of student arguments that we expect to see, based on the research-based strategies we have incorporated \cite{rebello2019scaffolding, uruena2017impact, rebello2013transfer, rebello2013exploring}; we aim to capture all possible emergent themes at scale. 
Therefore, while supervised approaches could, in principle, be used to label and predict features of student arguments, doing so would require substantial manual annotation and would likely limit the analysis to predefined categories (e.g., binary classifications). Unsupervised techniques allow us to analyze large volumes of text data efficiently and discover meaningful structure without the extensive human labeling, thereby enabling a more exploratory and data-driven understanding of students’ written arguments.

\subsubsection{Topic Modeling}
Topic modeling \cite{lee1999learning, blei2003latent} is an unsupervised machine learning technique used to uncover hidden thematic structures within large collections of text. Rather than relying on labeled data, topic modeling algorithms infer topics based on patterns of word co-occurrence across documents \cite{blei2012probabilistic}.

One widely used method of topic modeling is Non-negative Matrix Factorization (NMF) \cite{lee1999learning}, which we employed in this study. NMF is rooted in linear algebra and works by approximating an original document-term matrix (e.g. student essay) as the product of two lower-dimensional matrices. The key constraint in NMF is that all values must be non-negative, which tends to produce more interpretable results—especially useful when analyzing human language, where negative values lack intuitive meaning.

A simple explanation of NMF is as follows: suppose we have a document-term matrix $V \in \mathbb{R}^{m \times n}_{+} $, where $m$ is the number of documents and $n$ is the number of unique terms. NMF factorizes $V$ into two non-negative matrices: $W \in \mathbb{R}^{m \times k}_{+}$ and $H \in \mathbb{R}^{k \times n}_{+}$ such that their product approximates the original matrix:
\[
V \approx WH
\]
Where $k$ represents the number of topics. Each row in $W$ represents a document’s distribution over topics.
$H$ represents  a topic's distribution over the terms. 

In summary:
\[
W_{m \times k} \times H_{k \times n} = V_{m \times n}
\]
where $m$ is the number of documents, $k$  is the number of topics, and $n$ is the number of unique terms.

\section{Methods}

\subsection{Dataset}
This study was implemented in a first-semester calculus-based physics course for future engineers, at a large U.S. Midwestern land grant University. The annual enrollment in the course at the time of the study  is approximately 2500 students (1100 in fall and 1400 in spring). The course is built around three key principles: momentum, energy, and angular momentum, and it follows Chabay and Sherwood's \textit{Matter \& Interactions} \cite{chabay2011matter}. The course has three components: lectures (two 50-minute-long session), lab (one 110-minute-long session, and recitations (one 50-minute-long session). The recitations are the context of this study.

The recitations are led by one graduate teaching assistant (GTA) with the aid of an undergraduate teaching assistants (UTA). During each Recitation session, the GTA's spend a short time at the beginning of the recitation introducing the problem and relevant information. They start off each section by going over a provided Powerpoint presentation introducing the recitation problem. Each presentation takes around 5 to 10 minutes. In earlier weeks of the semesters, students receive more help on both the conceptual aspects of the problem (i.e. identifying the relevant principles and concept) as well as on the procedural aspects (i.e. selecting the appropriate system and surroundings). Students also receive scaffolding on the argumentation aspects of the problem, which is described below in more detail. As the semester progresses students continue to receive help on the conceptual aspects of the problem, but less help on starting the procedural aspects of it. Then students are expected to work together in groups of 3-5 at their table to solve the problem. They often use a whiteboard to create a collaborative workspace to discuss and share their work. As students work together, the GTA and UTA are available and walking around the room to help groups when they have questions. Even though they work in groups, every student is expected to provide their answers to questions in an individual Jupyter Notebook file (.ipynb extension). They edit this file on Google Colab in class. Students discuss the questions collaboratively and enter their answers to these questions and input an image of their written work. 
    
After they are finished, they are expected to submit the file and a PDF of the file into the Learning Management System (Brightspace). The notebook file is uploaded so we have access to easily extract student responses. A PDF is uploaded for grading purposes. Students are graded as a group by GTAs on a rubric provided. GTAs choose one student randomly from each group, grade their uploaded recitation file based on a rubric, and assign everyone in the group this grade. Students are not expected to spend much, if any, time on the recitation outside of the 50-minute section, so there should be very little individual work on the recitation other than expressing their answers in their own words. Due to the nature of the grading, students are encouraged to ask the TAs questions and make sure everyone in the group is on the same page before they leave the recitation session.

\begin{figure*}
    \centering
    \includegraphics[width=0.7\linewidth]{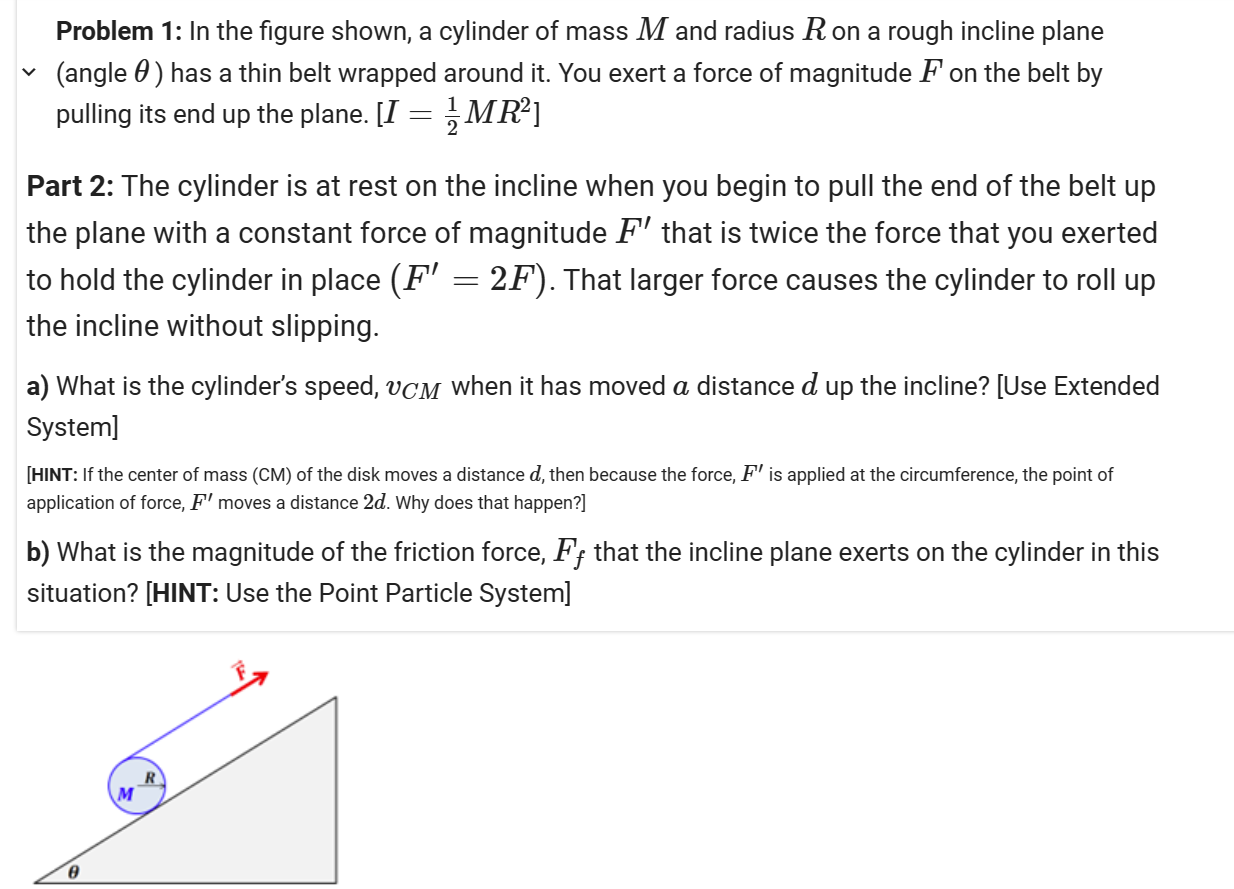}
    \caption{\centering Recitation Problem for Student Argumentation}
    \label{fig:R14_p2_sp23}
\end{figure*}

\subsection{Scaffolding Scientific Argumentation}
\begin{table*}[t]
    \caption{Summary of recitation scaffolds by semester.}
    \label{tab:scaffolding}
    \centering
    \setlength{\tabcolsep}{4pt} 
    \renewcommand{\arraystretch}{1.2} 
    \begin{tabular}{lccc}
        \hline\hline
        \textbf{Scaffolding} 
        & \textbf{Fall 2022} 
        & \textbf{Spring 2023} 
        & \textbf{Fall 2023 / Spring 2024} \\
        \hline

        \textbf{Module 1} &
        \parbox[t]{0.26\textwidth}{%
            Students were asked to construct arguments.
        } &
        \parbox[t]{0.26\textwidth}{%
            Students were given definitions of CER and asked to list their CER separately.
        } &
        \parbox[t]{0.26\textwidth}{%
            Students were given definitions of CER and prompts to identify CER statements.
        } \\
        
        \textbf{Module 2} &
        \parbox[t]{0.26\textwidth}{%
            Students were asked to construct arguments.
        } &
        \parbox[t]{0.26\textwidth}{%
            Students were given definitions of CER and asked to construct arguments.
        } &
        \parbox[t]{0.26\textwidth}{%
            Students were given definitions of CER and asked to list their CER separately.
        } \\
        
        \textbf{Module 3} &
        \parbox[t]{0.26\textwidth}{%
            Students were asked to construct arguments.
        } &
        \parbox[t]{0.26\textwidth}{%
            Students were given definitions of CER and asked to construct arguments.
        } &
        \parbox[t]{0.26\textwidth}{%
            Students were given definitions of CER and asked to construct arguments.
        } \\
        
        \hline
        \textbf{Study Question} &
        \parbox[t]{0.26\textwidth}{%
            In words, construct an argument to explain and justify your solution. Justify the various decisions you took while constructing your solution. In your argument, incorporate why the chosen principle(s)/concept(s) and assumption(s)/approximation(s) are relevant to your proposed solution.
        } &
        \parbox[t]{0.26\textwidth}{%
            In words, construct an argument to explain, elaborate and justify your solution. Your argument should be in a paragraph and contain the CLAIMS, EVIDENCE, and REASONING that support your solution.
        } &
        \parbox[t]{0.26\textwidth}{%
            In words, construct an argument to explain, elaborate and justify your solution. Your argument should be in a paragraph and contain the CLAIMS, EVIDENCE, and REASONING that support your solution.
        } \\
        \hline\hline
    \end{tabular}
\end{table*}

Throughout the last four semesters (Fall 2022 through Spring 2024), students were tasked with writing argumentation essays at the end of most recitations. Table \ref{tab:scaffolding} outlines how the scaffolds progressed each semester. The example question in the second row of each column represents the question students were asked at the end of the semester after receiving their full set of scaffolds. The dataset we report on is in response to question in Fig. \ref{fig:R14_p2_sp23}. 

The study started in Fall 2022 when students received no instruction on how to construct an argument; they were given the prompt shown in Table \ref{tab:scaffolding}. In Spring 2023, we implemented argumentation prompts based upon McNeil and Krajcik's model \cite{mcneill2010scientific} within the recitation. We gave students the definitions of claim, evidence, and reasoning (CER):

\begin{itemize}
    \item A CLAIM is every decision you took to solve the problem i.e. choice of principle/concept, system/surrounding, approximation/assumption, or key steps you used in your problem-solving strategy – It is an assertion that that the decision was the right one to answer the key question(s) in the problem statement.
    \item EVIDENCE is the information either provided in the problem or something that you know (e.g. from the physics class) to be factually correct that helped you make the decision above.
    \item REASONING is an explanation of why the EVIDENCE supports the CLAIM.
\end{itemize}

We provided the definitions starting with claims as the semester progressed, while prompting students to list their own CER in parts before writing a full argument of their own. Finally in Fall 2023 and Spring 2024 students were given similar scaffolds to that of Spring 2023. However, instead of immediately having students list their own CER, we gave students statements and had students label which statements were claim, evidence, or reasoning. Then in the middle portion of the semester students listed their own CER statements. Finally they constructed their own arguments at the end of the semester. The breakdown of these scaffolds and the final argumentation question asked to students are listed in Table \ref{tab:scaffolding}.

\subsection{Topic Modeling}

We analyzed open-ended student scientific arguments from  each of the four semesters separately. Although, students were given instructions to use only words and complete sentences to construct their essays, the data still required significant cleaning of the text. We thus began by employing a comprehensive text pre-processing pipeline to clean student argumentation essays. This included punctuation and number removal, conversion to lowercase, and removal of stopwords (e.g. 'the', 'and' , 'as', etc.) filtering using the Natural Language Toolkit (NLTK) \cite{bird2009natural}. Additionally, we implemented automated spell checking using the PySpellChecker \cite{pyspellchecker} library to correct common typos. Responses with fewer than ten words were filtered out to reduce noise.

We applied Non-negative Matrix Factorization (NMF) to extract latent topics from the student essays.  First, the pre-processed text was vectorized using the TF-IDF method, capturing term importance across the corpus. We then applied the NMF algorithm \cite{lee1999learning} with ten components (topics). Each component represents a distribution over words, and each essay is represented as a mixture of topics. 

 One common method to determine the optimal number of topics is known as the "elbow method" where you plot a model fit, such as reconstruction error\cite{lee1999learning} or coherence scores \cite{coherence2015}, across a number of topics. To identify the optimal number of topics for NMF, we plotted both the reconstruction error \cite{lee1999learning} and coherence scores \cite{coherence2015} across topic counts ranging from two to thirty. The reconstruction error, which measures the discrepancy between the original TF-IDF matrix and its reconstruction from the NMF model, was calculated for each topic configuration. This metric is used to assess how well the NMF model approximates the original data. Though the reconstruction error decreased monotonically with additional topics reflecting better matrix approximation a definitive "elbow" did not emerge for any semester. For the coherence scores, which measure the semantic interpretability of topics, fluctuated and did not exhibit a clear maximum. Since there was no clear quantitative metric to identify ideal number of topics, we selected 10 topics in line with prior work \cite{bralin2023analysis} using similar data context.

\subsection{Interpretation of Topics}
 To compare the semesters, we analyzed the representative words of each topic, the number of essays in each topic, and the overall distribution of student responses across topics using a concentration analysis \cite{bao2001concentration} of each semester. To interpret the topics across semesters, we examined the top-weighted words associated with each component from the NMF model. Each essay was assigned to the topic with the highest topic weight. We then visualized topic distribution across the corpus using histograms that highlight the number of student essays per topic. 
 
 The concentration analysis introduced by Bao and Redish \cite{bao2001concentration} was originally developed to measure how students' responses on multiple-choice questions are distributed. Since we are interested in measuring student membership to a topic, this analysis will apply a novel insight into our student distribution. In this study, we have applied this method to quantify how student essays are distributed over different topics. The concentration factor, $C$, is calculated as: 

 \[C= \frac{\sqrt{m}}{\sqrt{m}-1} \times \frac{\sum_{i=1}^{m}n_i^2}{N} - \frac{1}{\sqrt{m}}\]

Where, $m$ represents the number of topics, $n_i$ represents the total number of students who selected topic $i$, and $N$ represents the total number of students in the semester. A high concentration factor $C \approx 1$ means most essays are concentrated in one or a few topics, where as $C \approx 0$ suggests an even spread across all topics.

\section{Results}
\begin{figure*}[t]
\centering
    \begin{subfigure}[b]{0.45\textwidth}
        \centering
        \includegraphics[width=\textwidth]{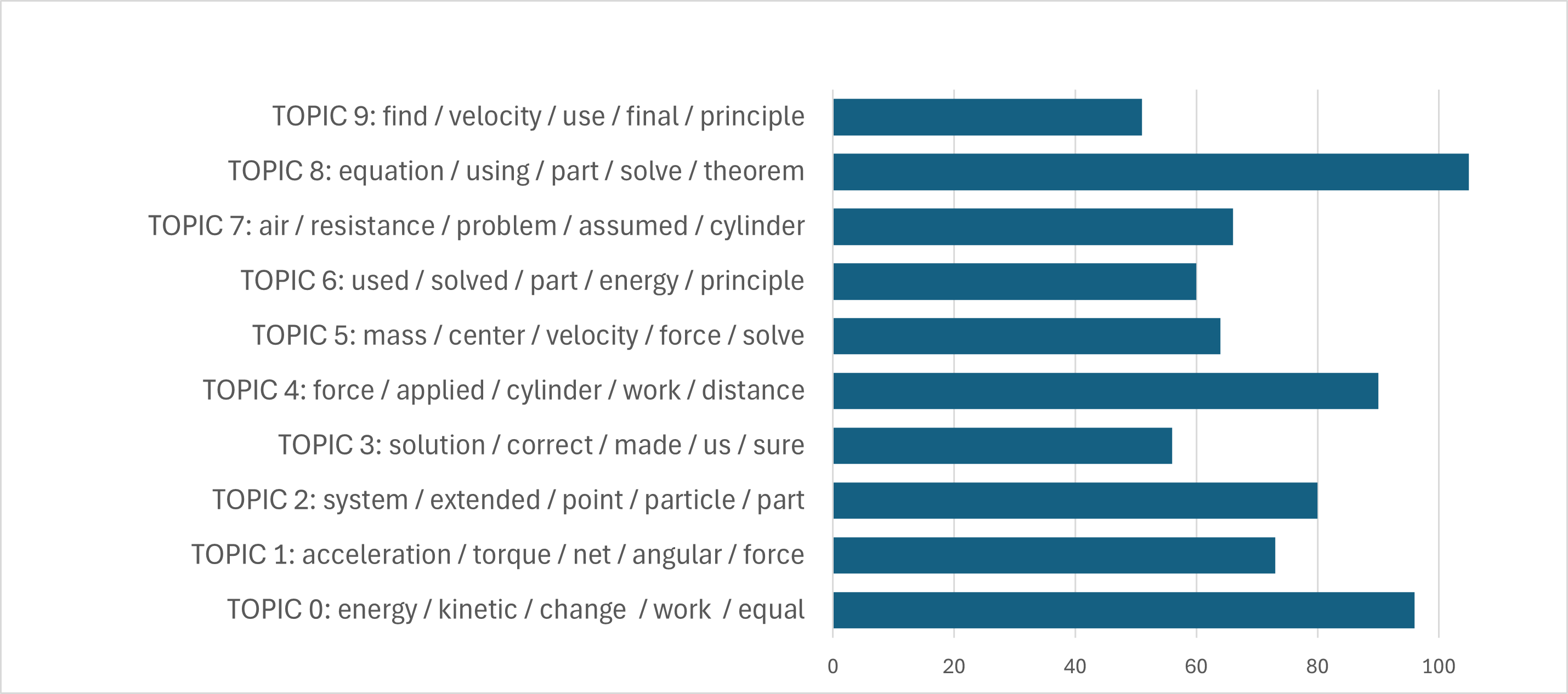}
        \caption{\centering Fall 2022}
        \label{fig:Num_Essays_NMF_F22}
    \end{subfigure}
    \hfill
    \begin{subfigure}[b]{0.45\textwidth}
        \centering
        \includegraphics[width=\textwidth]{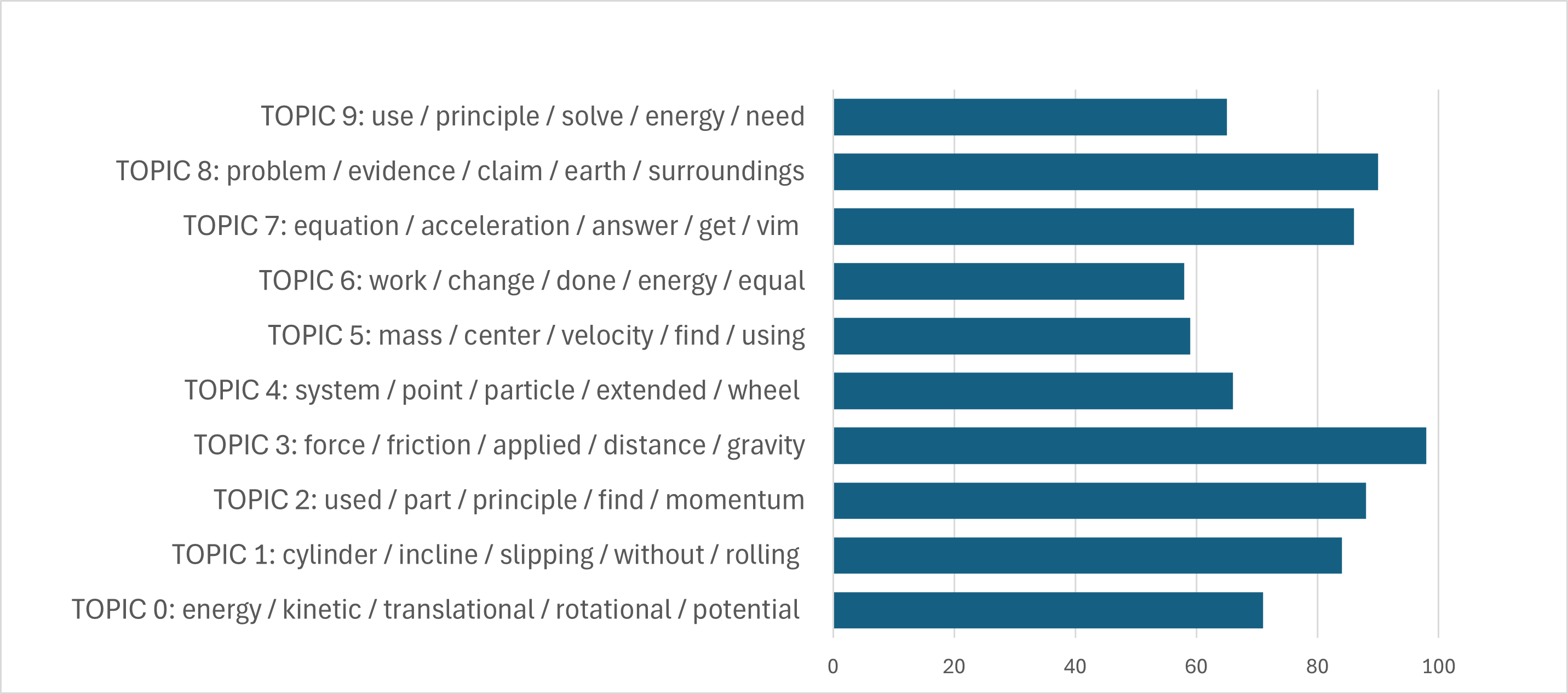}
        \caption{\centering Spring 2023}
        \label{fig:Num_Essays_NMF_SP23}
    \end{subfigure}
    
    \vspace{1em}
    
    \begin{subfigure}[b]{0.45\textwidth}
        \centering
        \includegraphics[width=\textwidth]{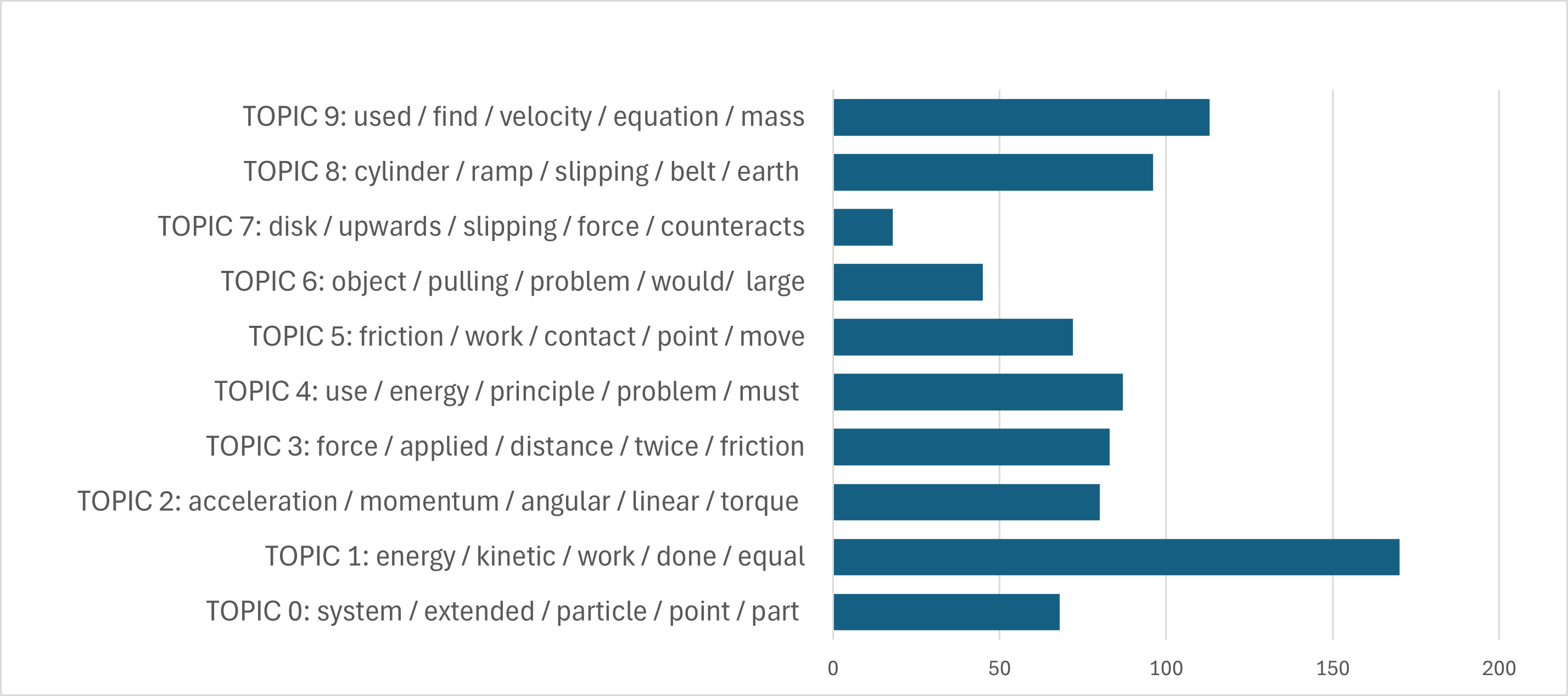}
        \caption{\centering Fall 2023}
        \label{fig:Num_Essays_NMF_F23}
    \end{subfigure}
    \hfill
    \begin{subfigure}[b]{0.45\textwidth}
        \centering
        \includegraphics[width=\textwidth]{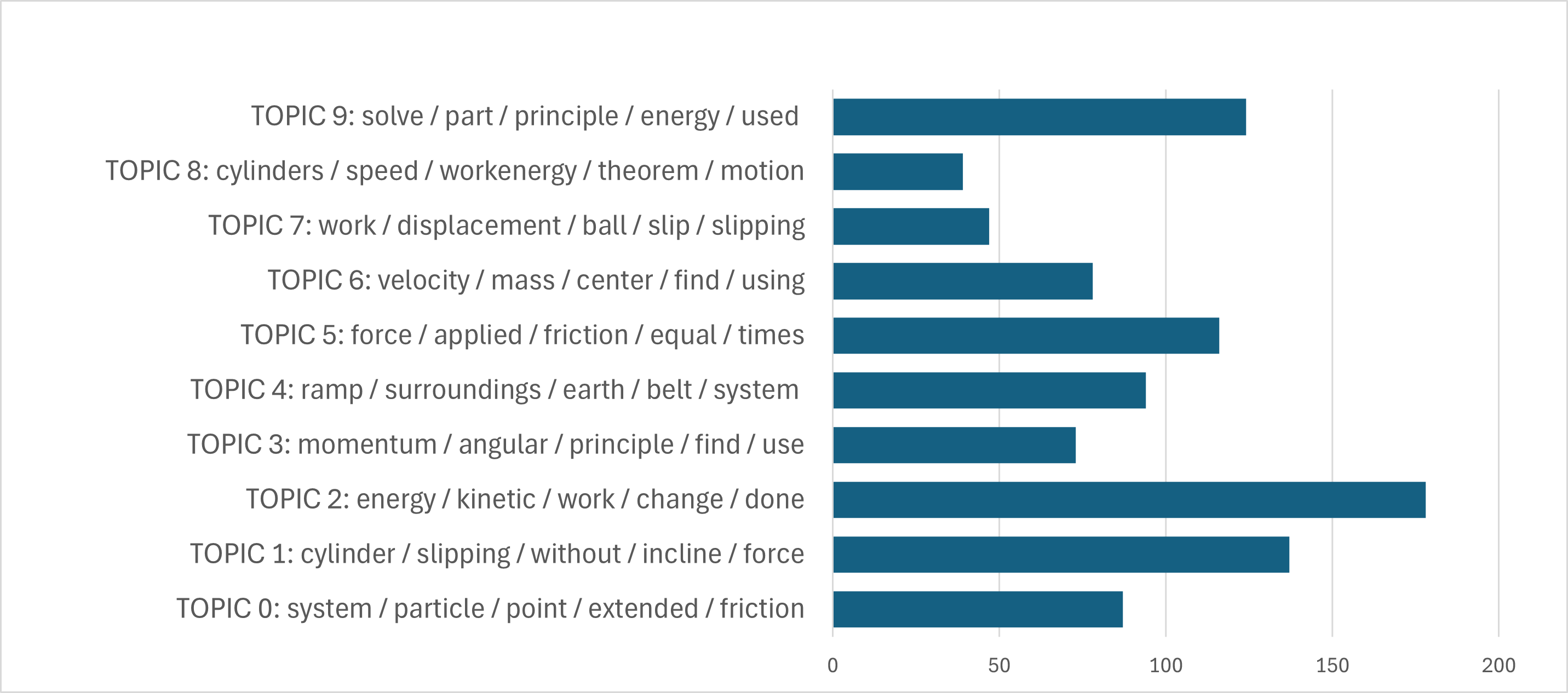}
        \caption{\centering Spring 2024}
        \label{fig:Num_Essays_NMF_SP24}
    \end{subfigure}
    
    \caption{\centering Representative words and distribution of essays in NMF topics across scaffolded semesters}
    \label{fig:Num_Essays_NMF_All}
\end{figure*}

Our goal is to utilize topic modeling to determine if we can capture the longitudinal change in the broad themes of student arguments throughout the semester. To achieve this we report on the top five words of each topic, exemplar essays from relevant topics, and the essay distribution over topics of each semester. 

The resulting top five words were calculated to assess what each topic represented, shown in Figure \ref{fig:Num_Essays_NMF_All}. The number of essays in each topic are represented in the bar graphs of Figures \ref{fig:Num_Essays_NMF_F22}, \ref{fig:Num_Essays_NMF_SP23}, \ref{fig:Num_Essays_NMF_F23}, and \ref{fig:Num_Essays_NMF_SP24} shown below. We strove to improve students argumentation with the scaffolds in such a way that students focused more on the underlying physics principles of the problem as opposed to the surface features or mathematical steps of solving the problem. For this specific problem, a correct solution would focus on using the Energy Principle: $\Delta E_{system}=W_{surroundings}$ and concepts involving energy. Therefore, we would ideally see the top words of the topics focused on types of energy and/or work. 

First, we looked at Fall 2022 topic distribution and top words in Figure \ref{fig:Num_Essays_NMF_F22}. We found the topics to be fairly evenly distributed with minor peaks at Topics 0, 4, and 8. Our overall takeaway from these representative words is that there was some focus on explicit physics principles (i.e. Topics 0 and 4), which became apparent from emergent words such as "energy", "kinetic", and "force". There was much more focus on mathematical steps (i.e. Topics 3, 6, 7, 8, and 9) shown from the representative words, such as "solution", "used", "equation" and "find". To further look into the topics we report on a representative essay from the topic with the highest student membership. For Fall 2022, this would be Topic 8. Each student essay is comprised of variations of topics. For an essay to be categorized in a certain topic the essay must have a majority percentage of that topic. In Topic 8, for example, one of the most representative essays is classified as 84.41 \% of Topic 8, 11.03 \% of Topic 2, and less than 5 \% of other topics. The student wrote \textit{"By setting the system to everything we can make the equation equal to zero. From here, all we need to do is identify everything in the equation, plug in the relating equations, and simplify."} This student does not discuss any physics principles to any degree and focuses on their mathematical steps, which is in agreement with our classification of Topic 8 consisting of essays focusing on procedure.

Second, we looked at Spring 2023 in Figure \ref{fig:Num_Essays_NMF_SP23}. We found the topics to be fairly evenly distributed with slight valleys at Topics 4, 5, 6, and 9. Our overall takeaway from these representative words is that there was some focus on explicit physics principles (i.e. Topics 0, 3, 4, and 6), shown from emergent representative words such as "energy", "force", "particle", and "work". An exemplar essay from a student belonging to Topic 3 is \textit{Friction force is reliant on the normal force. Because the weight doesn't change and an additional force is not applied, the normal force does not change.}. Though the student is not focusing on energy, they are clearly thinking about other physics involved in solving the problem, such as friction. The other topics  there was much more focus on mathematical steps (i.e. Topics 2, 5, 7, and 9), which we see from the representative words such as "used", "equation", and "solve", and surface features (i.e. Topic 1), which we see from the representative words such as "cylinder" and "incline".

Third, we looked at Fall 2023, shown in Figure \ref{fig:Num_Essays_NMF_F23}. We found the topics to be not as evenly distributed with a large peak emerging at Topic 1. Our overall takeaway from these representative words is that there was some focus on the energy physics principle (i.e. Topics 0, 1, and 4), as words such as "energy", "kinetic", and "extended" are present amongst the top words of these topics. Note that the topic that has the most student membership is Topic 1 which focused primarily on energy. The student essay most representative of Topic 1 is \textit{"the total work of the system is equal to translational kinetic evergy plus rotational kinetic energy. This is beacause no other types of energy were gained or lost during this phase. This means we can equal net work to the sum of the energies."} This is an exemplar student essay as the student constructs an argument centered around energy principles, such as relating total work to the relevant energy, and conservation of energy. Other topics within this semester that focused on other physics concepts were Topics 2, 3, and 5 where words such as "momentum", "friction", and "force" appear. In addition, the other topics focused on mathematical/procedural steps (i.e. Topic 9), evident from words such as "used" and "find", and surface features (i.e. Topics 6, 7, and 8) concluded from words such as "cylinder", "disk", and "object".

Finally, we looked at Spring 2024 in Figure \ref{fig:Num_Essays_NMF_SP24}. We found the topics to be not as evenly distributed with a large peak emerging at Topic 2. Our overall takeaway from the representative words is that there was some focus on the correct physics principle (i.e. Topics 0 and 2) where words such as "particle" and "energy" emerged. Note that the topic that hand the most student membership is Topic 2 which we tagged as focusing primarily on energy due to the representative words. The student essay that was most representative of Topic 2 was \textit{"We used the energy principle to in which the change in energy is equal to the work done by the surroundings, work, which is equal to the net force times d. We are informed by the problem that there will be a change in translational kinetic energy and no change in rotational kinetic energy. Since we are informed of that there is a change in translational kinetic energy and no change in rotational kinetic energy we can set change in Ktrans equal to the work done. After identifying the forces acting on the system and summing them, we get the net force times the displacement on the right side of the equation and the change in translational kinetic energy on the left side. We plug in our representative variables and then solve for the Vcm."}. Similar to the Fall 2023 student response, this student focused on energy principles as well, shown by their discussion of the relationship between work and energy. This response also contained elements (though significantly less) of Topic 9 which focused in part on mathematical steps and energy. Other topics that focused on other physics concepts were Topics 3, 5 and 6 illustrated by the representative words, such as "momentum", "velocity", and "force". There were still topics that focused on mathematical steps (i.e. Topic 9), with words like  "solve" and "part" emerging, surface features (i.e. Topics 1 and 4), shown by words such as "cylinder" and "ramp" , and a mixture of these surface features, mathematical steps, and physics principles (i.e. Topics 5, 6, 7, and 8).

\begin{figure}
    \centering
    \includegraphics[width=0.75\linewidth]{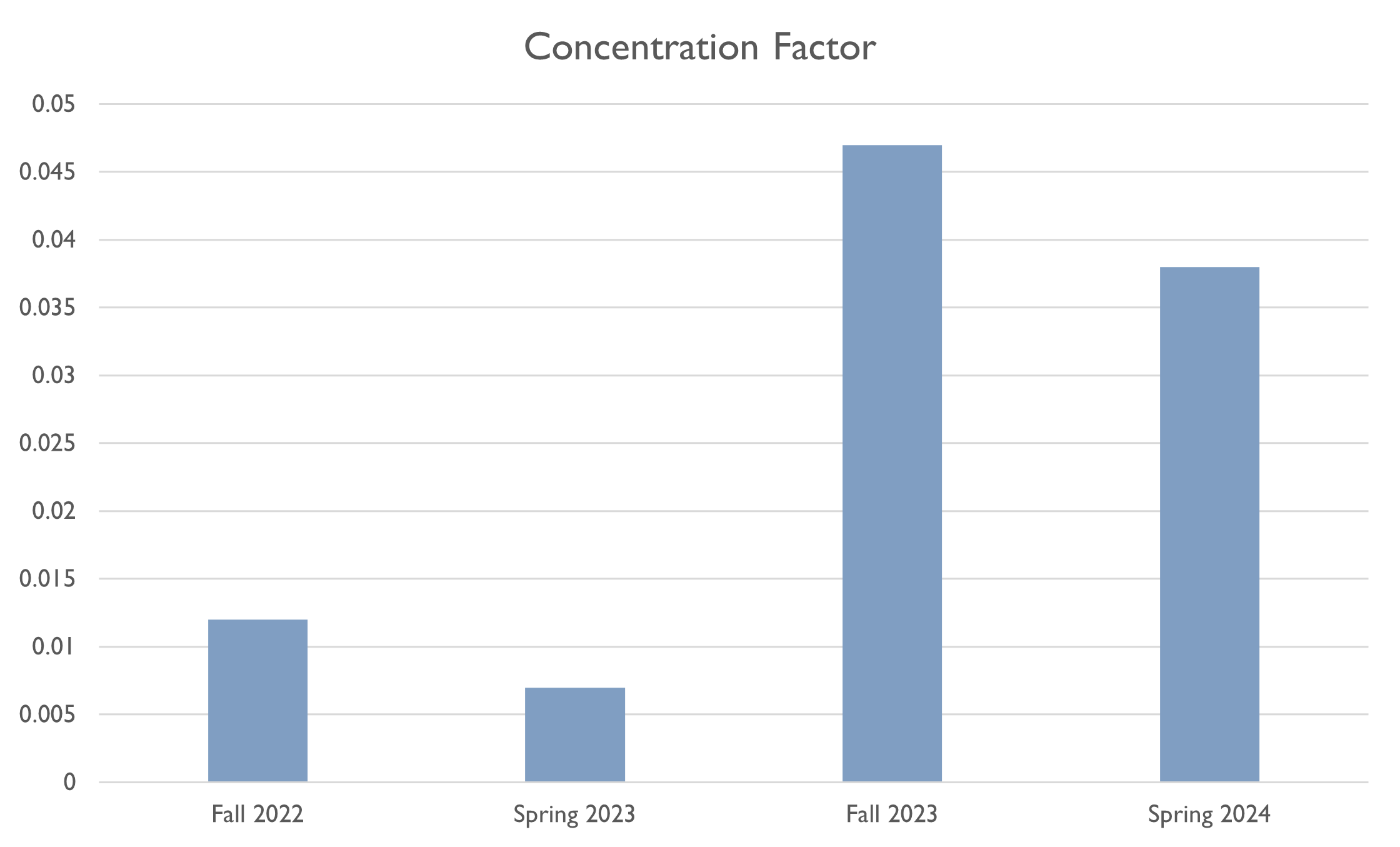}
    \caption{\centering Concentration Factors over Four Semesters}
    \label{fig:concentration}
\end{figure}

To establish statistical significance of the peaks, we calculated Bao and Redish's concentration factor \cite{bao2001concentration} for each semester, shown in Figure \ref{fig:concentration}. From this calculation, we see that the Fall 2022 and Spring 2023 semesters have a much lower concentration than that of semesters Fall 2023 and Spring 2024 (4$\times$ larger). Interestingly, we also see a small dip in concentration between fall and spring semesters. From Fall 2022 to Spring 2023 the concentration factor decreases, and the same happens from Fall 2023 to Spring 2024. Overall, this analysis supports what we see from the distribution. The Fall 2022 and Spring 2023 semesters have less of a concentration in a particular topic than that of Fall 2023 and Spring 2024.

\section{Discussion}

After investigating the top words and topic distribution, we observed that in Fall 2022 and Spring 2023, during which minimal argumentation scaffolding was provided, the topics seemed to be evenly distributed. These results suggest that in general students focused equally on relevant concepts as much as on trivial concepts, such as mathematical steps. 

In Fall 2023 and Spring 2024, in which students received more extensive scaffolding, we saw unequal distributions, with peaks in topics that emphasized physics concepts (e.g. energy). This trend indicates that students in the later semesters tended to focus more on physics-oriented ideas. Ideally, students should focus on the underlying physical principles while constructing arguments, as this reflects conceptual reasoning. Conversely, focusing primarily on mathematical steps reflects an emphasis on procedural aspects of problem-solving.

To further investigate the distribution across topics, we conducted a concentration analysis and report on the concentration factor in each semester in Figure \ref{fig:concentration}. Overall, we found an increase in concentration from Fall 2022 to Spring 2024.  The Fall 2023 and Spring 2024 semesters, in which students received the most training in argumentation, have higher concentrations of essays than Fall 2022 and Spring 2023 semesters. According to Sadler et al. \cite{sadler2016understanding}, when students achieve a more consistent understanding of the underlying concepts, their responses tend to exhibit less variability. Therefore, the observed increase in concentration may reflect more conceptually coherent and consistent reasoning among students in the later semesters.

Taken together, these findings support two key outcomes. First, the increase in concentration indicates that students' arguments focus more on the topical physics principles rather than procedural details. Second, since these concentration factors were calculated based upon essay membership across topics, this result also demonstrates that topic modeling successfully captured broad shifts in students’ argumentation patterns across semesters.

It is also interesting to note an emergent pattern between fall and spring semesters. The concentration of Spring 2023 is less than that of Fall 2022, and the concentration of Spring 2024 is less than that of Fall 2023. This recurring pattern suggests a potential difference between students enrolled in fall versus spring semesters. Specifically, students in the fall semesters may demonstrate a stronger ability to construct arguments grounded in physics principles. 

One possible explanation lies in the known differences of student populations in fall and spring semesters. Students enrolled in the fall tend to enter the course with a stronger physics and calculus background from high school coursework. In contrast, students in the spring generally have less prior physics experience and are more likely to be required to take a prerequisite calculus course.

\begin{figure}
    \centering
    \includegraphics[width=0.75\linewidth]{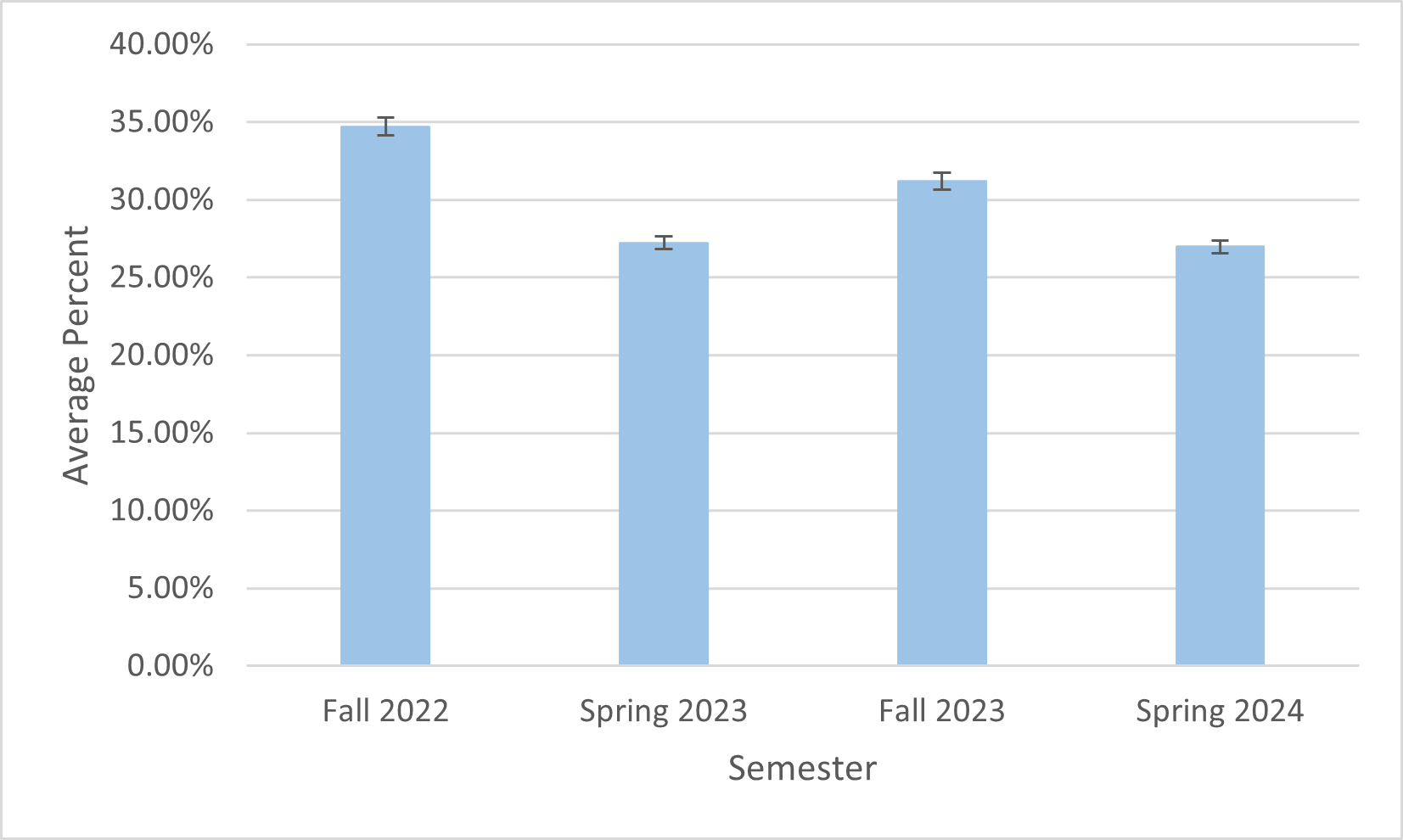}
    \caption{\centering EMCS Average Scores at Start of the Semester}
    \label{fig:EMCS_pre_scores}
\end{figure}

To investigate if this pattern extends beyond the topic modeling results, we examined scores from the Energy and Momentum Conceptual Survey (EMCS) \cite{singh2003} administered at the beginning of each semester in the laboratory section of the course. We calculated the average scores for each group of students, shown in Figure \ref{fig:EMCS_pre_scores}. The same pattern emerged: students in the spring semesters scored lower on average on the EMCS pre-test than those in the fall semesters. This indicates that students in the Spring semesters had lower prior knowledge than those in the Fall semesters, which is consistent with a greater variability in argumentation quality in the spring semester as shown by the lower concentration factor in the spring semester argumentation data. 

This convergence between topic modeling results and independent conceptual measures lends further validity to the concentration factor as an indicator of conceptual focus. Moreover, it suggests that the observed differences between semesters are not merely artifacts of the topic modeling process, but may instead reflect genuine differences in students’ prior preparation and conceptual understanding.


\section{Conclusions}
In summary, using ML, we analyzed the top words, distribution of topics, and the concentration factor of the topics obtained from NMF. Through this analysis and across all student responses, we were able to identify topics that reflected on energy concepts, mathematical steps, surface features, etc. that all matched the context of the given problem. To address our research question: \textit{How does the quality of students’ scientific argumentation in the context of physics problem-solving evolve in response to scaffolding and fading over multiple weeks of instruction?} we examined the top words and distribution of topics over four semesters of varying argumentation scaffolds to detect any changes.  In semesters with minimal scaffolding (Fall 2022 and Spring 2023), students’ arguments were distributed more evenly across topics, indicating an equal emphasis on both conceptual and procedural elements. In contrast, in semesters with greater scaffolding (Fall 2023 and Spring 2024), topics were more concentrated around conceptually rich ideas, particularly those emphasizing energy principles.These results suggest that explicit argumentation scaffolds can help students shift from procedural reasoning toward more conceptually grounded argumentation in physics. 

By investigating both the top words and topic distribution across semesters, we observed a clear change in student focus as instructional scaffolds were implemented. In the earlier semesters, student focus was distributed equally across conceptual and procedural topics. In the later semesters, we observed their focus to concentrate around physics principles, such as conservation of energy, associated with the problem statement.  Overall, these findings indicate that structured scaffolding in scientific argumentation instruction supports students in constructing more conceptually meaningful explanations over time. These findings demonstrate that unsupervised ML methods, such as topic modeling, can capture large-scale trends in student argumentation, offering a scalable and reproducible approach for characterizing conceptual versus procedural focus in written physics responses.
\\

\subsection{Limitations, Implications, and Future Work}

The results of this study highlight the potential of unsupervised machine learning methods as tools for evaluating student written argumentation at scale. While promising, this approach is limited by the interpretive nature of topic modeling and the need for human validation to ensure meaningful topic labeling. Nevertheless, these methods can serve as efficient diagnostic tools for instructors seeking to track changes in student reasoning across semesters. There were also inherent limitations related to the implementation of scaffolds within the course. First, scientific argumentation was taught and scaffolded only within the recitation section of the course; students did not receive any formal instruction on scientific argumentation from lecture or lab. This restricted the extent of student exposure and may have limited opportunities for reinforcement. Second, since scaffolds occurred exclusively in recitation, the quality and depth of instruction was partly reliant upon GTA discretion. Although GTAs received weekly training and were  provided with standardized slides introducing scientific argumentation, the extent to which they discussed these concepts varied. Nevertheless, all students received identical written scaffolds through the recitation worksheets, which helped maintain a baseline level of consistency in the instructional material.

Future work will expand this analysis to track individual students’ scientific arguments longitudinally within a single semester to explore whether similar trends in conceptual focus can be detected over shorter timescales. We also plan to extend the dataset to include additional semesters to examine whether the observed differences between fall and spring cohorts persist. More broadly, this line of work points toward the development of machine learning–based frameworks that can provide instructors with scalable, evidence-based feedback on student scientific argumentation in physics and beyond.

\section{Acknowledgments}

This research is supported in part by U.S. National Science Foundation grants 2111138. Any opinions, results, and findings expressed here are those of the authors and not of the Foundation.

\bibliography{references}
\clearpage
\appendix
\onecolumngrid

\appendix
\onecolumngrid

\end{document}